\documentclass[aps, twocolumn,noshowpacs,preprintnumbers,amsmath,amssymb]{revtex4}

\usepackage{verbatim}
\usepackage{wasysym}

\usepackage{graphicx}
\usepackage{dcolumn}
\usepackage{bm}
\usepackage{wrapfig}

\begin{document}

\title{Tailoring the diameter of decorated C–N nanotubes by temperature variations using HF-CVD}

\author{Ralph Kurt}
\author{Christian Klinke}
\author{Jean-Marc Bonard}
\author{Klaus Kern}
\author{Ayatollah Karimi}
\affiliation{Departement de Physique, Ecole Polytechnique Federale de Lausanne, CH-1015 Lausanne EPFL, Switzerland}

\begin{abstract} % abstract goes here

Patterned films of decorated nitrogenated carbon (C–N) nanotubes were catalytically synthesised by hot filament chemical vapour deposition (HF-CVD) in a nitrogen–methane–ammonia environment. The systematic study of a transition between different kinds of C–N nanostructures as a function of the local substrate temperature ranging from 700 up to 820$^{\circ}$C is presented. The morphology, the diameter as well as the properties of the generated tubular structures showed strong dependence on this parameter. By means of electron microscopy a new type of decoration covering all tubular structures was observed. Buckled lattice fringes revealed the disordered graphitic-like character of the hollow C–N nanotubes. Raman spectroscopy confirmed a transition in the microscopic order as a function of temperature. Furthermore field emission in vacuum was studied and showed a spectacular correlation to the deposition temperature and therefore the diameter of the
C–N tubes. For arrays of tubes thinner than 50 nm an onset field below 4 V/$\mu$m was observed.

\end{abstract}

\maketitle

\section*{Introduction}

The discovery of fullerene C$_{60}$ by Kroto et al. in 1985 \cite{1} has sparked a lot of exciting research in the field of carbon based nanostructures. Numerous new structures like carbon nanotubes \cite{2}, nano-onions \cite{3}, nano-horns \cite{4} or micro-trees \cite{5} were reported. Most of these materials hold a variety of interesting properties including electronic \cite{6,7} and mechanical properties \cite{8,9}, which promise to be not just of academic interest but also to be sources for new applications. As an example carbon single and multiwalled nanotubes have been used as cold field emitters (e.g. \cite{10}) and as hydrogen storage material \cite{11}. 

On the other hand, carbon nitride (CN$_{x}$) coatings and diamond like carbon (DLC) films have been under intense investigation for a longer time because they can serve as hard layers for tools (e.g. \cite{12}). Interesting mechanical behaviour, in particular an exceptionally high elasticity, has also been observed for sputtered CN$_{x}$ thin solid films \cite{13,14}. In that case, the strengthening effect has been attributed to a cross-linked fullerene-like structure with curved and buckled basal planes \cite{13}.

\begin{figure}[ht]
  \centering
  \includegraphics[width=0.45\textwidth]{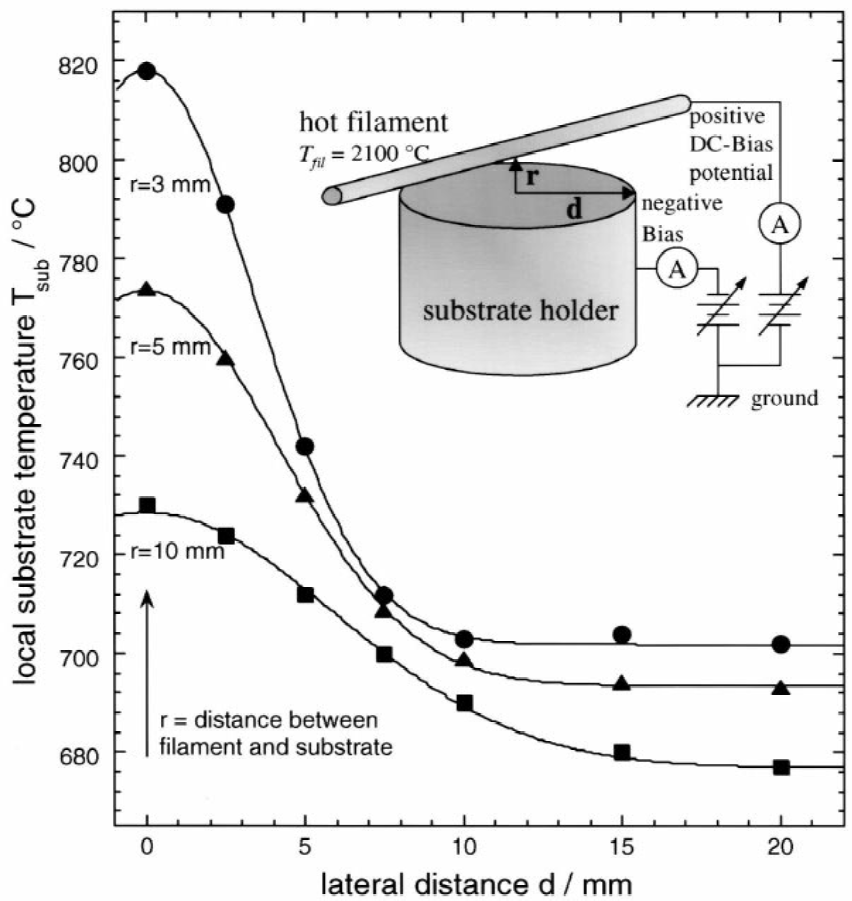}
  \caption{\textit{The measured temperature profiles (filled symbols) at the sample surface versus the lateral distance $d$ from the filament were fitted (calculated lines) assuming a Gaussian temperature distribution. Due to thermal radiation from the filament (T$_{fil}$ = 2100$^{\circ}$C) a stronger gradient is observed for smaller distances $r$ between filament and substrate. The inset gives a schematic drawing of the deposition configuration consisting of a resistively heated filament placed close to the substrate holder.}}
\end{figure}

Furthermore, doping of carbon nanotubes with nitrogen or boron is expected to induce exciting electronic properties \cite{15}. Miyamoto et al. \cite{16} showed that tubular forms of graphitic-like carbon nitrides can be stabilised by displacing C atoms out of the honeycomb network of graphite. Detailed transmission electron microscopy (TEM) and electron energy loss spectroscopy (EELS) of carbon nitride nanotubulites prepared by chemical vapour deposition (CVD) and magnetron sputtering, 
respectively, was reported by Suenaga et al. \cite{17,18}. Several other groups have also succeeded in synthesising nitrogen-containing nanotubes \cite{19,20,21,22,23}. Furthermore, nanotubes within the boron nitride (BN) \cite{23,24} and ternary B-C-N systems \cite{22,25,26} have been reported. Apart from the structure, both the electronic and mechanical properties are dependent on the chemistry of hetero-atomic nanotubes (e.g. \cite{27}). However, the correlation between the morphology, crystalline order and properties of such tubular CN$_{x}$ structures is not yet completely understood. 

Recently we described the autocatalytic growth of a new type of carbon structures \cite{28} produced by hot filament chemical vapour deposition (HF-CVD). They consist of tubular nitrogenated carbon (C-N) nanostructures covered with graphitic-like sheets growing perpendicular to the tube axis. Furthermore the essential role of nitrogen on structure formation of decorated nanotubes was analysed in detail by TEM and EELS \cite{29}. At this stage various decorated nanotube-like structures were grown randomly distributed at the surface of a pure Si substrate, and factors that influence nucleation were not understood. Vapour-phase deposition of CN$_{x}$ nanotubes with controlled structure and chemical composition can provide a tool for synthesis of materials with tailored electronic \cite{16} and mechanical properties \cite{30}.

\begin{figure*}[ht]
  \centering
  \includegraphics[width=0.9\textwidth]{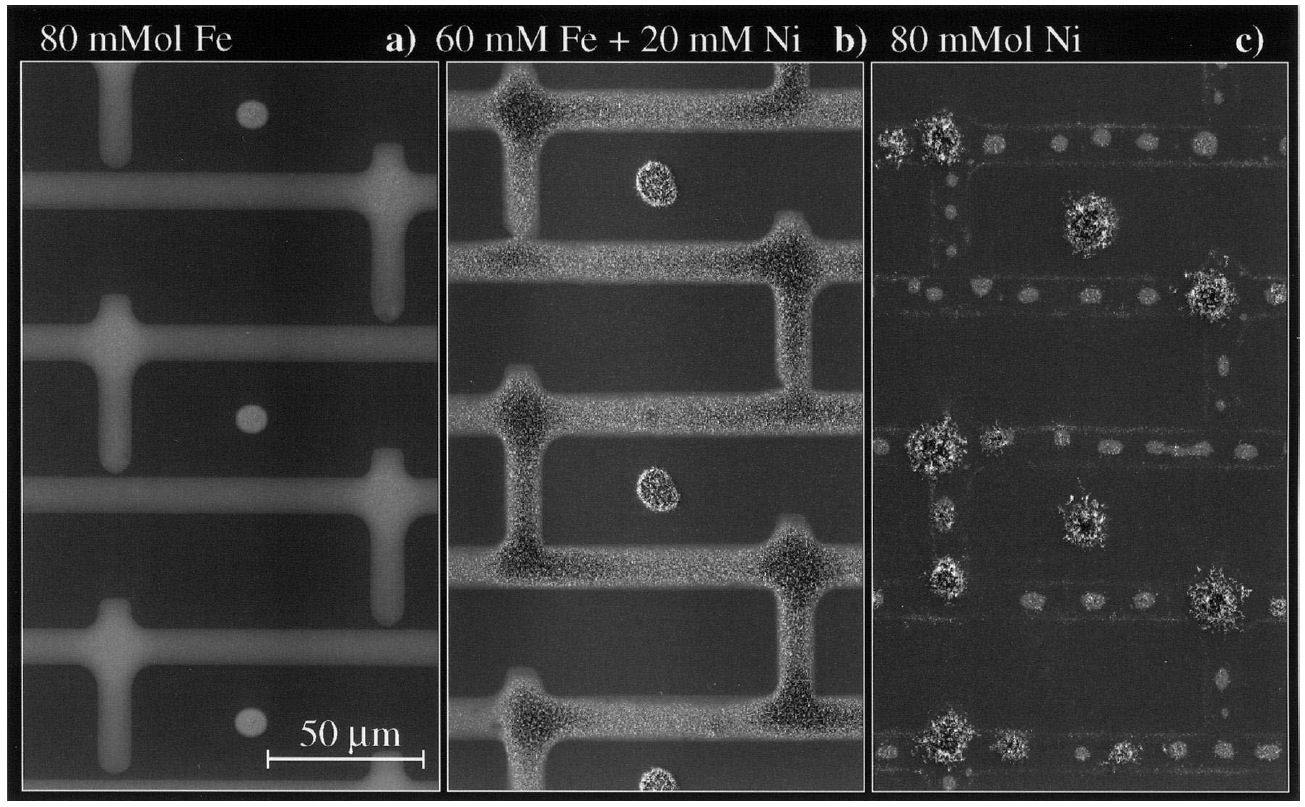}
  \caption{\textit{Scanning electron microscopy images of patterned films as grown by HF-CVD at a lateral distance $d$ = 5 mm within 30 min ($r$ = 3 mm). Growth rate and printing behaviour strongly depend on the type and concentration of the used catalyst.}}
\end{figure*}

We present here a systematic study of the transition between different kinds of C-N nanostructures as a function of temperature (given by lateral distance from filament), which determines the tube diameter. Strong correlation to properties such as Raman signal and field emission were found. The fundament of these investigations was the deposition of patterned films of decorated nanotubes, which were realised by applying a catalyst on the substrate using the microcontact ($\mu$CP) printing method.

\section*{Experimental methods}

\textit{Synthesis of nanostructured material}

Microcontact printing ($\mu$CP) of catalysts \cite{31} is becoming a popular method because it allows to form chemical patterns on the surface of a variety of substrates. Structured stamps were realised by curing the elastomer poly(dimthyl)siloxane (PDMS) over a master. They were then hydrophilised by an oxygen plasma treatment and stored under water to keep them hydrophilic. The stamps were dried under nitrogen flow for 10 s before use. The catalyst ink solution (Fe(NO$_{3}$)$_{3}$ $\cdot$ 9H$_{2}$O and Ni(NO$_{3}$)$_{2}$ $\cdot$ 6H$_{2}$O solved in ethanol) was applied to the stamps after a period of 12 h of ageing, which 
was found to be ideal for catalytic growth of nanotubes \cite{32}. The printing was done by bringing the inked stamp in contact with the surface of a Si wafer for 3 s. 

Decomposition of methane (CH$_{4}$) using the bias-enhanced hot filament chemical vapour deposition technique results in a new type carbon based nanostructures. Pure 
nitrogen was used as dilution gas and a small flow of ammonia (NH$_{3}$) was introduced during the experiments in order to increase the amount of nitrogen in the deposits. The total gas flow was kept constant at 500 sccm and a typical flow ratio of N2-CH4-NH3 of 100:1:1 was applied.

\begin{figure*}[ht]
  \centering
  \includegraphics[width=0.9\textwidth]{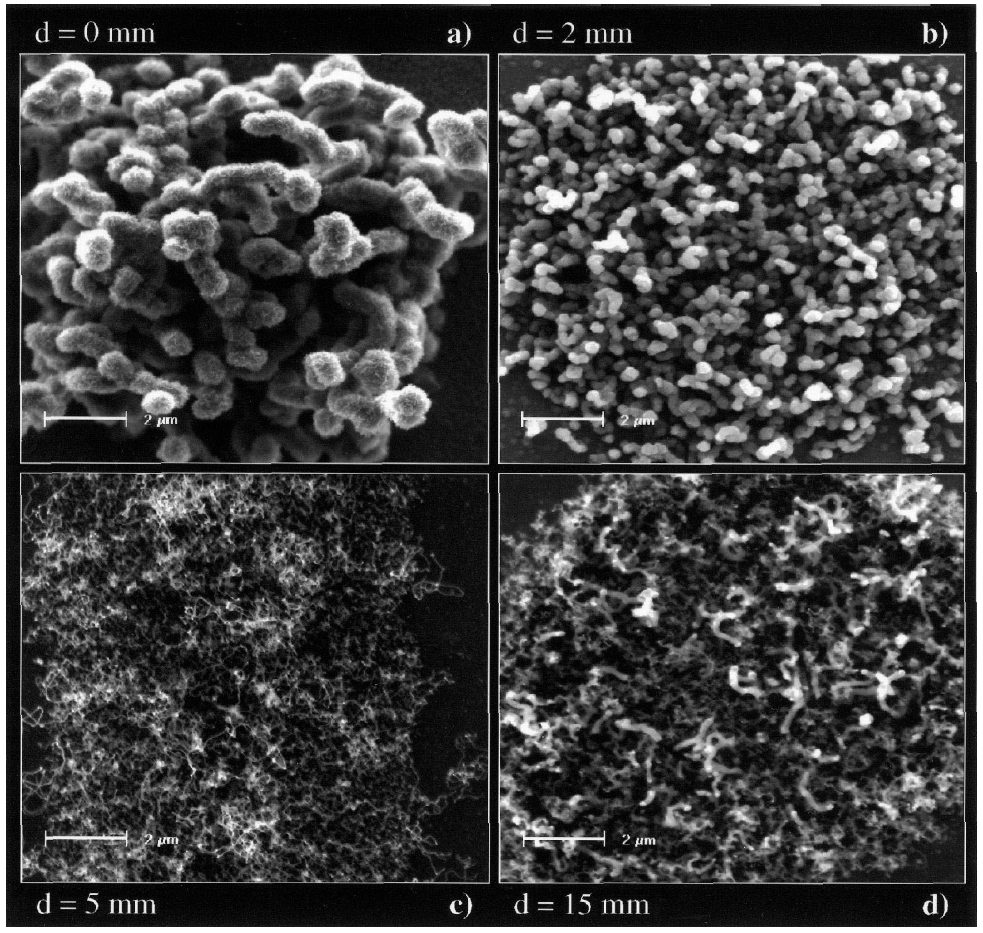}
  \caption{\textit{SEM plan view micrographs showing the evolution of shape and diameter of wounded nanotubes as a function of the lateral distance $d$ ($r$ = 3 mm). The surface of each tubule prepared by HF-CVD is always covered with thin lamellas, resulting in a coral-like structure. For all images the same magnification is used.}}
\end{figure*}

For all the experiments described in this work, only one resistively heated tungsten carbide (WC) filament ($\diameter$ = 500 $\mu$m) was used in order to study the resulting nonuniformity in deposition conditions described by the lateral distance $d$ at the sample surface measured perpendicular to the filament. The inset in Fig. 1 shows schematically the experimental setup. A molybdenum (Mo) substrate holder placed at $r$ = 3 mm (fixed value for all further reported results) underneath the filament was heated independently to 650$^{\circ}$C. Positive and negative dc-bias potential up to 200 V were applied independently to the filament and to the substrate, respectively. The background pressure typically reached in the deposition chamber was 10$^{-5}$ Pa, whereas the working pressure $p$ was kept constant at 300 Pa during deposition. A fixed filament temperature T$_{fil}$ = 2100$^{\circ}$C as measured by means of a two-colour pyrometer was used. The temperature profile at the sample surface (T$_{sub}$ versus $d$) as measured by a W-5\%Re/W-26\%Re thermocouple is given in Fig. 1. Due to the thermal radiation a strong temperature gradient was observed depending on the distance between filament and substrate ($r$). The measurement data were fitted assuming a Gaussian distribution. 

\text{} \\
\textit{Characterisation techniques}

Scanning electron microscopy (SEM) was performed to analyse the microstructures in plan view. A Philips XL 30 microscope equipped with a field emission gun (FEG) 
operating at an acceleration voltage between 2 and 5 kV, a working distance of typically 10 mm, and in secondary electron (SE) image mode was used. 

The growth morphology of the tubular structures and their crystallinity were controlled by analytical transmission electron microscopy (TEM). For this purpose a 
Hitachi HF-2000 field emission microscope equipped with a Gatan image plate operating at 200 kV (point resolution 0.23 nm) was used.

\begin{figure}[ht]
  \centering
  \includegraphics[width=0.45\textwidth]{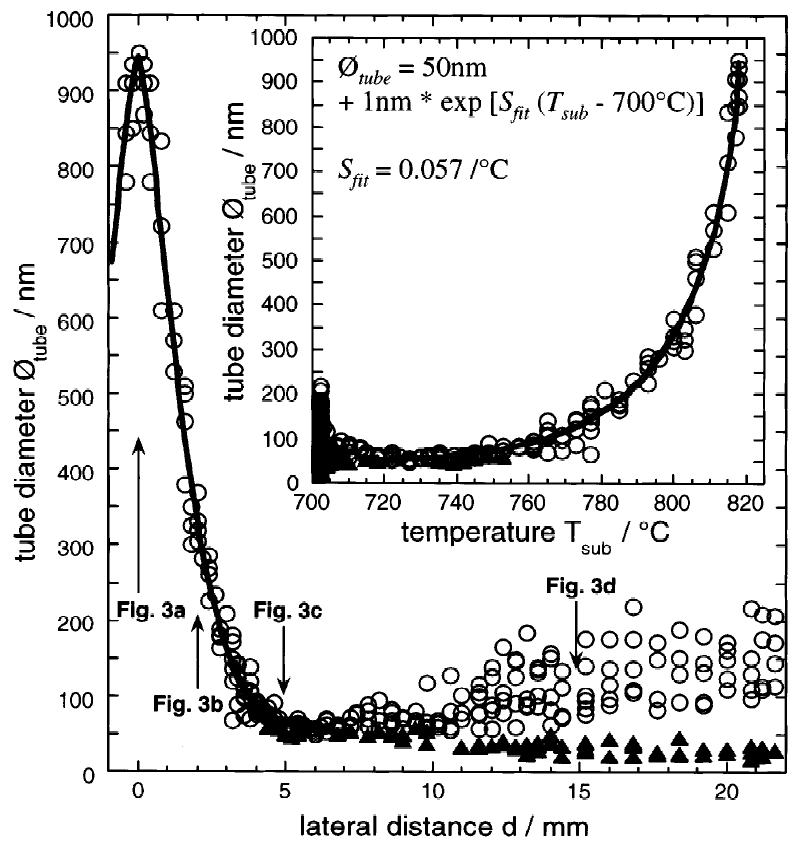}
  \caption{\textit{The diameter of the tubular structures as obtained from SEM shows a strong dependence on the lateral distance $d$ ($r$ = 3 mm). Maximum diameters were observed directly underneath the filament. For $d$ $>$ 5 mm a large dispersion of the tube diameters was found (open and filled symbols). The positions where Figures 3a-d were taken from are marked. The inset shows the tube diameter versus the local substrate temperature.}}
\end{figure}

The micro Raman technique was used to obtain local information about the vibrational properties of the nanostructures. The Raman spectra were recorded in 
backscattering configuration using the 514.5 nm line of an Ar$^{+}$ ion laser and a DILOR XY 800 spectrometer. An incident maximal laser power of 20 mW was applied in order to avoid peak shifts due to thermal heating during data acquisition or structure transformations. A spot size of approximately 10 $\mu$m was achieved with a 100$\times$ Olympus microscope objective. The spectra were calibrated using a natural diamond single crystal. 

Additionally, field emission measurements were performed using the examined samples as cathodes with a lateral resolution of $\sim$1 mm. The emitted electrons were 
collected on a highly polished stainless steel spherical counterelectrode of 1 cm diameter, which corresponds to an emission area of $\sim$0.007 cm$^{2}$. The distance was initially adjusted to 125 $\mu$m and was reduced to 50 or 25 $\mu$m when no emission was observed below 1000 V. A Keithley 237 source-measure unit was used for sourcing the voltage (up to 1000 V) and measuring the current with pA sensitivity, allowing the characterisation of current-voltage (I-V) behaviour.

\section*{Results}

\textit{Morphology}

We used $\mu$CP to study the specific effect of the catalyst on both the quality of the printed pattern and the tube growth. Figure 2 shows SEM images of the substrate after deposition using various catalyst mixtures. The structure consists of lines and dots with a typical structure width of 10 $\mu$m. Best printing was achieved with an Fe catalyst as shown in Fig. 2(a) whereas the pure Ni solution showed poor printing behaviour. The Ni ink produced small drops on the stamp of $\sim$10 µm diameter, which were transferred to the substrate as seen in Fig. 2(c). In contrast, a higher efficiency regarding the growth of nanotubes was obtained for Ni than for pure Fe. The best results in terms of both tube growth and patterns uniformity were achieved using an ethanolic ink containing 60 mMol Fe$^{3+}$ and 20 mMol Ni$^{2+}$. This ink was used for the following experiments. Note that the same type of decorated tubes also grows autocatalytically \cite{28} with the disadvantage of a random distribution on the surface of a pure Si wafer. Clearly, a well-controlled film growth requires the use of a catalysts.

\begin{figure*}[ht]
  \centering
  \includegraphics[width=0.9\textwidth]{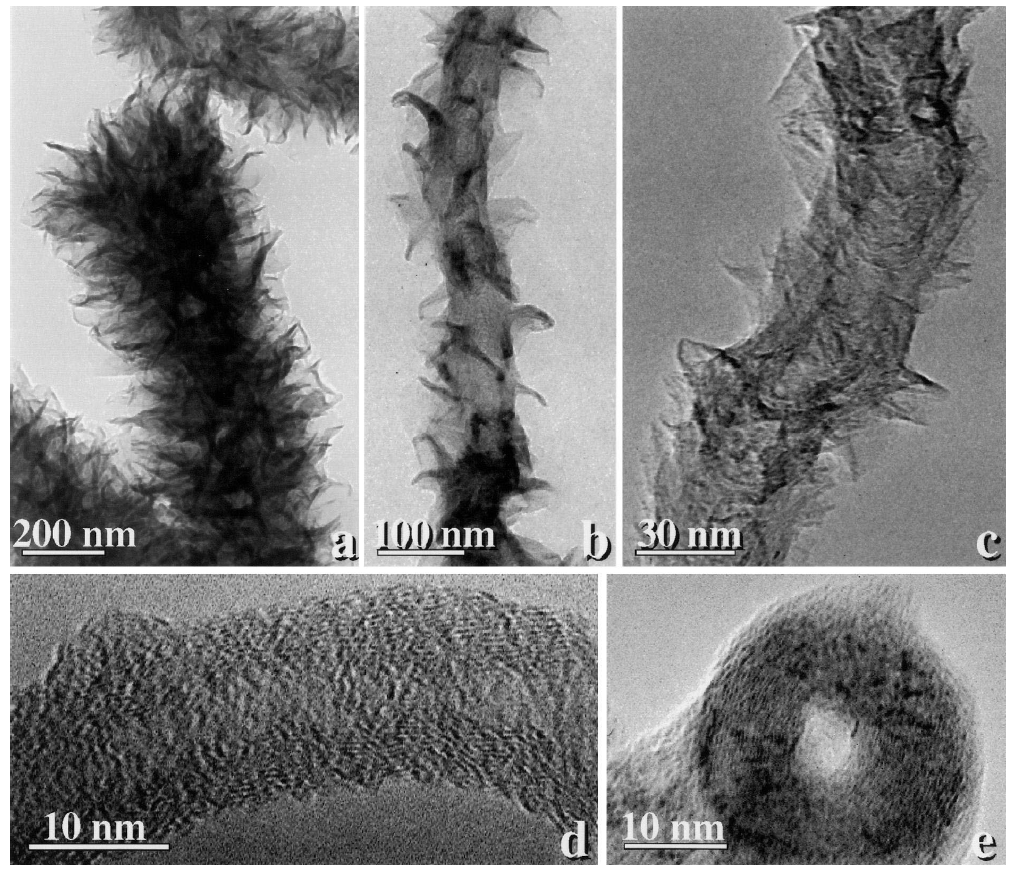}
  \caption{\textit{TEM analysis of nitrogenated carbon nanotubes with different tube diameter and degree of coverage. The decoration consists of thin graphitic-like foils growing perpendicular to the tube axis. High resolution plan view (d) and cross section (e) reveals disordered graphitic-like structure typically characterised by buckling of the graphene planes due to the incorporation of nitrogen. The contrast indicates hollow tubes and a rough tube surface.}}
\end{figure*}

The above mentioned temperature gradient T$_{sub}$ versus the lateral distance from the filament $d$ (Fig. 1) strongly affects the growth conditions for the nitrogenated carbon tubes. The evolution of the morphology of various tubular structures grown onto the substrate in dependence of this important parameter is represented in Fig. 3 by a series of SEM images. For Figures 3(a) to 3(d) the lateral distance d increases from 0 to 2, 5, and 10 mm, respectively. From $d$ = 0 to 5 mm (Figs. 3(a-c)) the structures become finer, i.e. the tube diameter decreases from approximately 1 $\mu$m to 50 nm. Above $d$ = 10 mm the local substrate temperature remains constant at T$_{sub}$ $\approx$ 700$^{\circ}$C (Fig. 1). Simultaneously a significant change in the morphology of the grown tubes is 
observed. As one can conclude from Fig. 3(d) nanotubes with a strong dispersion in diameter (20 to 200 nm) were formed at the same location, i.e. under the same 
conditions. In contrast, for higher T$_{sub}$ ($d$ $<$ 7 mm) a sharper local distribution of tubular C-N species in terms of tube diameter was obtained (Fig. 3(a-c)). 

As a rule, all structures are of wounded tubular character. It is interesting to note that a decoration is formed over the surface of all studied tubes, which can be clearly seen in Fig. 3a. The large dendritic-like surface of the decorated C-N tubes results from thin foils or lamellae growing perpendicular to the tube axis. The shape of these nanoaggregates varies with the applied filament temperature, ranging from twisted lamellae to straight sheets showing ramifications \cite{33}. Recently, we could show \cite{29} that coiling as well as decoration depends on the amount of nitrogen incorporated in the tubes. Performing energy electron loss spectroscopy (EELS) a maximum N/C atomic ratio of 0.043 $\pm$ 0.014 was obtained on samples produced under the same conditions but without catalyst \cite{34}.

\begin{figure}[ht]
  \centering
  \includegraphics[width=0.45\textwidth]{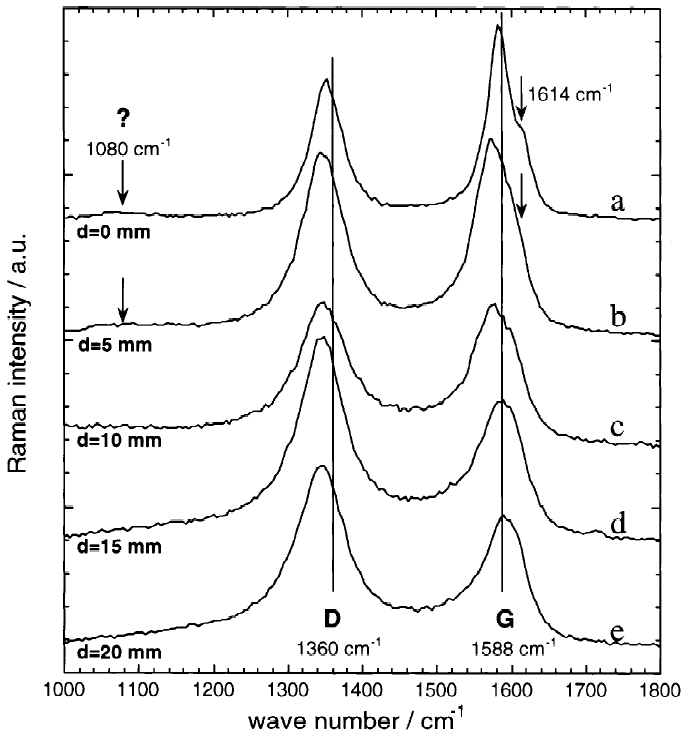}
  \caption{\textit{Comparison of Raman spectra obtained from decorated C-N nanotubes grown at different lateral distances $d$ increasing from 0 to 20 mm. Beside the graphite G peak and the disordered graphitic D peak a shoulder at 1614 cm$^{-1}$ and an unidentified peak at 1080 cm$^{-1}$ were detected. In dependence on the tube diameter as well the peak ratio I$_{G}$/I$_{D}$ changes as peak shifts regarding to the nominal vibration frequencies (vertical lines) were observed.}}
\end{figure}

A quantitative analysis of the tube diameter ($\diameter _{tube}$) was done using high-magnification SEM. Figure 4 shows the observed data of the outer tube diameter as a function of the lateral distance $d$. It confirms the strong dependence on $d$ and therefore on the local substrate temperature. The diameter was found to vary between 15 and 950 nm. Maximum tube diameters of approximately 1 $\mu$m were observed directly underneath the filament. For $d$ $<$ 5 mm we found that the larger $d$, the thinner the diameter and the longer the tubes. Wounded tubes with a length up to 50 $\mu$m were observed with typical diameters of approximately 50 nm. 

For $d$ $>$ 7 mm the distribution of the tube diameter becomes broader with increasing $d$. In Figure 4 the thinnest tubes were marked by filled symbols whereas open symbols were used for the thicker species. The positions where Figures 3a-d were taken are also marked in Fig. 4. The inset shows the tube diameter ($\diameter _{tube}$) versus the local substrate temperature (T$_{sub}$). In the analysed region one can fit an exponential curve $\diameter _{tube} = 50 \cdot \exp \left[ S ( T_{sub} - 700^{\circ}C ) \right]$ nm to the obtained data, where $S$ is a fit parameter related to the slope of the tube diameter. The values 50 nm and 700$^{\circ}$C are offsets for the tube diameter and the temperature, respectively. $S$ was determined to be 0.057/$^{\circ}$C. Considering the nice fit of the calculated curve to the experimental data, it is obvious that the nanotube growth depends exponentially on the local temperature. The higher the deposition temperature the thicker the decorated C-N nanotubes. The same behaviour was observed using other catalyst concentrations and even for the autocatalytic growth. 

TEM investigations were performed to study the microstructure of decorated nanotubes in more detail. Figure 5 shows typical micrographs of such C-N nanotubes with decreasing outer diameter obtained at increasing $d$. The same tendency as given in Fig. 4 was confirmed by TEM analysis. Furthermore Figs. 5(a-c) reveal again  structures radiating from the surface of the nanotubes. High resolution TEM images suggest that the needle-like structures are in fact thin bent or rolled up graphene sheets.

\begin{figure}[ht]
  \centering
  \includegraphics[width=0.45\textwidth]{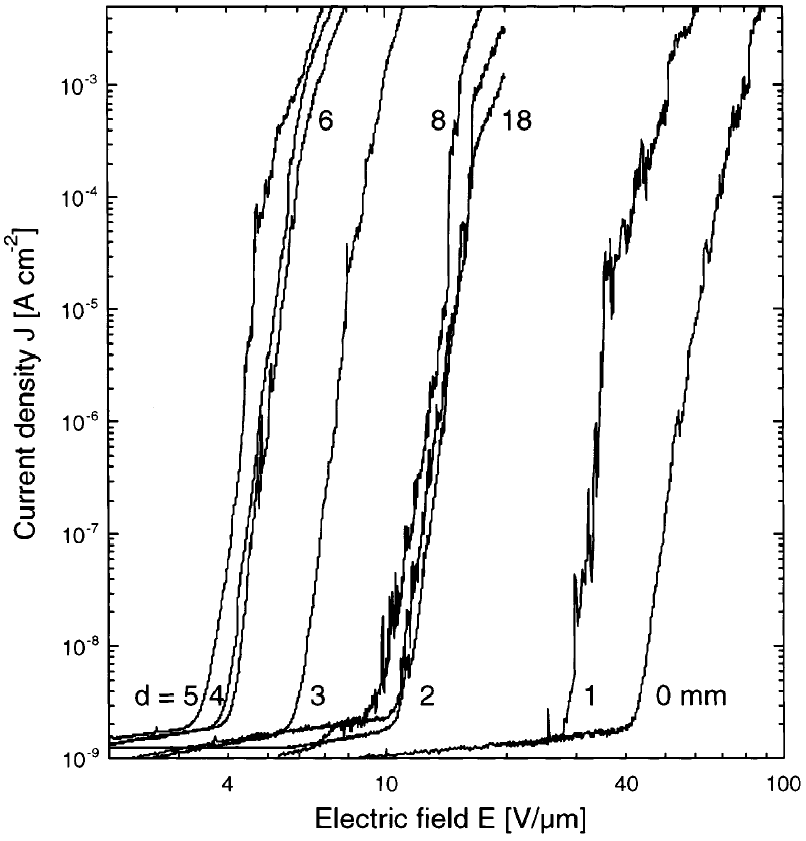}
  \caption{\textit{Current density $J$ versus applied electric field $E$ for patterned films of C-N nanotubes. The turn-on E$_{to}$ and threshold fields E$_{thr}$ required to extract a current density of 10 $\mu$A cm$^{-2}$ and 10 mA cm$^{-2}$, respectively, first decrease before increase again as a function of the lateral distance $d$ (as indicated). Lowest values E$_{to}$ = 4.6 V/$\mu$m and E$_{thr}$ = 7.2 V/$\mu$m were observed at approximately $d$ = 5 mm. From a mathematical fit using the Fowler-Nordheim model, field amplification factor $\beta$ of 750 was determined in that case.}}
\end{figure}

In the case of thick nanotubes the inner structure is difficult to study due to the decoration that reduces the electron transparency. The contrast in Figs. 5(d-e) clearly represents the tubular structure of very thin C-N nanotubes and suggests that they are probably hollow inside. Tubes aligned parallel to the electron axis (Fig. 5(e)) indicate concentric wall structures and an inner diameter in that case of less than 5 nm. The two high magnification micrographs of individual tubes in plan view (Fig. 5(d)) and cross section (Fig. 5(e)) reveal the occurrence of a crystalline structure. The lattice fringes visible in Figs. 5(d,e) correspond to the \{002\} basal planes of graphite. Locally the planes are roughly parallel to each other but on a larger scale they show bending and interlinking between planes. Such a buckling effect of the basal planes reflecting the disorder even on atomic scale has been also observed in previously reported CN$_{x}$ structures \cite{14,17,18} but also in pure carbon nanotubes catalytically grown \cite{35}.

\text{} \\
\textit{Raman spectroscopy}

The influence of the local substrate temperature T$_{sub}$ during deposition of the various types of C-N tubular structures was investigated by Raman spectroscopy. The bonding states in carbon based materials depend on the nature of polymorphs, which is influenced by the deposition parameter. These effects are illustrated in Fig. 6, which compares the micro-Raman spectra generated from nanotubes grown at different lateral distances d increasing from 0 to 20 mm as indicated. 

All spectra were decomposed into at least two strong peaks confirming the graphitic-like structure of the material. It is known that pyrolythic graphite leads to a sharp vibration mode at 1588 cm$^{-1}$ \cite{36} due to the presence of C-sp$^{2}$ domains (first order G band). Structural modifications result in a shift of this peak to lower wave numbers as seen in curves (a-c). A shift to higher wave numbers, i.e. higher activation energy of the vibration mode, as seen in curve (e) might be caused by internal stress in the material. A second strong peak at approximately 1360 cm$^{-1}$ is considered to represent a more disordered structure labelled as D (disordered) band \cite{37}. Decreasing particle size, bending of the lattice fringes as well as incorporation of N into the structure may activate this band and cause furthermore the observed peak-shifts. Note that in a perfect graphitic crystal this first order vibrational mode is forbidden due to the selection rules. 

Analysing the G band in more detail, a shoulder located at approximately 1614 cm$^{-1}$ becomes clearly visible (marked by an arrow in Fig. 6). This additional vibration mode appears at high T$_{sub}$, i.e. $d$ between 0 and 4 mm and it is less pronounced below 750$^{\circ}$C. Beside that, a significant shift of the measured G peak position was observed as a function of $d$. The largest deviation $\Delta$ = 15 cm$^{-1}$ was detected at $d$ = 5 mm (curve b) corresponding to decorated C-N nanotubes with an outer diameter less than 50 nm. 

It is worth studying the relative Raman peak area intensities of the D and G band (I$_{D}$/I$_{G}$). It was found that the I$_{D}$/I$_{G}$ peak ratio increases from $\sim$0.5 for spectrum (a) to $\sim$2.5 (spectrum e). This behaviour can visibly verified in Fig. 6 since the D band becomes dominant for large $d$. In line with the discussion above, the D band arises from the finite size of the graphite domains and originates from grain boundaries or imperfections, such as substitutional N atoms, sp$^{3}$ carbon, or other impurities \cite{38}. Therefore the relative Raman peak area intensities of the D and G band are a measure of the degree of order of graphite clusters in the C-N tubes, which decreases with increasing $d$. According to \cite{39} one can estimate the length of the graphite cluster $L_{graphite}$ from the I$_{D}$/I$_{G}$ ratio using the following equation: $L_{graphite} = 4.4$ nm $\cdot (I_{D}/I_{G})^{-1}$. In the case of our decorated C-N nanotubes, graphite cluster sizes of approximately 9 and 2 nm, respectively, were obtained for $d$ = 0 and 20 mm. 

The half width of the peaks varies only slightly in the presented curves and reflects either the degree of crystalline perfection in the case of sharp peaks or the more amorphous-like character of the material (broader overlapping peaks as in curve (e)). An additional band located at approximately 1080 cm$^{-1}$ was detected at the tubular structures deposited at small lateral distances but it could not yet identified.

\text{} \\
\textit{Field emission}

Field emission is one of the most promising applications for carbon-based films, and recent studies have stressed the determinant influence of the film morphology on the emission properties \cite{10,40,41,42,43}. The ability to tailor the tube diameter by HF-CVD could thus prove to be a decisive advantage to produce emitting structures of high quality. 

The effect of the distance to the filament, and hence of the tube diameter, is displayed in Fig. 7. The field emission was measured at 20 locations on the film between $d$ = 0 and 20 mm as indicated, and Fig. 7 shows clearly a strong variation in the field emission properties. Directly underneath the filament ($d$ = 
1 mm), no field emission below 1100 V applied voltage (which corresponds to the limit of our acquisition setup) was observed unless the interelectrode distance was decreased to 10 $\mu$m. The onset field (applied field necessary to obtain a current density of 10 nA cm$^{-2}$) was 36 V/$\mu$m, and the threshold current density of 10 mA cm$^{-2}$ was not obtained below 95 V/$\mu$m. As $d$ increased, the fields decreased rapidly (see Fig. 7), and reached a minimum for $d$ = 5 mm. The interelectrode distance was 125 $\mu$m at that point, and the onset and threshold fields were 3.7 and 7.2 V/$\mu$m, respectively. Beyond that distance, the emission fields increased again, and levelled off for distances larger than $d$ = 8 mm. In that region, the interelectrode distance was 50 $\mu$m, and the onset and threshold 
fields amounted to $\sim$10 and $\sim$20 V/$\mu$m, respectively. 

Field emission results are usually analysed with Fowler-Nordheim theory. The Fowler-Nordheim model describes the electron emission from a flat surface under a 
high applied field by tunnelling through the triangular surface potential barrier. The emitted current, $I$, is proportional to $I \propto F^{2} \cdot \exp(B \phi ^{3/2} /F)$, where $F$ is the applied field just above the emitting surface, $\phi$ is the workfunction and $B$ is a constant ($B = 6.83 \cdot 10^{9}$ V eV$^{3/2}$ m$^{-1}$) \cite{44}. In the case of sharp point emitters such as the ones considered here, $F$ is generally not known and is therefore usually taken as $F = \beta E = \beta V / d_{0}$, with $V$ the applied voltage, $d_{0}$ the interelectrode distance and $E = V / d_{0}$ the macroscopic applied field. The field amplification factor $\beta$ is determined solely by the geometrical shape of the emitter \cite{45}. 

In the case of Fig. 7, all emission curves followed the Fowler-Nordheim model at low emitted currents (up to $\sim$10 $\mu$A cm$^{-2}$), and the field amplification factor could thus be easily determined by fitting the emission characteristics with the Fowler-Nordheim formula, using a workfunction of $\phi$ = 5 eV \cite{46}. The field amplification factor $\beta$ follows a behaviour which is reciprocal to that of the emission fields: the field amplification is very low just below the filament ($\beta$ = 80 at $d$ = 0 mm). It increases rapidly with the distance and reaches a maximal value of $\beta$ = 750 for the most efficient tubes, and decreases then again to $\beta$ $\sim$ 300 far away from the filament.

\section*{Discussions and Conclusions}
 
The fact that all our C-N nanotubes prepared by HF-CVD are always decorated may probably interpreted by at least two simultaneous growth mechanisms. The results 
allow one to suppose a tubular growth (along the tube axis) on the one hand and the formation of graphitic foils on the other hand. In the later case one can assume as a hypothesis that small flake-like aggregates are already formed in the plasma close to the filament and are attached to the outer surface of the tubes, where they can act as nuclei for a growth of the needle-like graphene sheets. The observed splitting of the Raman G peak at high T$_{sub}$ may probably also originated by one more ordered and another disordered graphitic-like structure corresponding to the two different growth mechanisms. 

In the temperature range from T$_{sub}$ = 700 to 820$^{\circ}$C a continuous increase of the tube diameter (50-1000 nm) with temperature was observed. This strong correlation seems to be a behaviour of universal character due to its independence on the major deposition parameter such as catalyst, kind and concentration of hydrocarbon gases, T$_{fil}$, $r$, bias potential and working pressure (over large ranges). 

Since several years pure carbon fibres have been grown from the decomposition of hydrocarbons (e.g. methane, acetylene, benzene, natural gas) at lower temperatures (e.g. $\sim$300$^{\circ}$C - 700$^{\circ}$C) in the presence of a metallic catalyst (e.g. Fe, Ni, Co) \cite{47}. The diameter of such carbon filaments range from several hundred micrometers to about 100 nm and decreases with increasing temperature. Sporadic works also showed synthesis of hollow tubes in this temperature region \cite{48}. Moreover thermal decomposition of acetylene in an oven at temperatures above 720$^{\circ}$C results as well in catalytic growth of nanotubes whose diameter increases with temperature \cite{49}. However, carbon based nanotubes seem to exhibit minimum diameter in the temperature range from 720 to 750$^{\circ}$C confirmed by the present work. 

The dispersion of the tube diameters above 7 mm may be attributed to nonequilibrium growth conditions in this region far from the filament. Due to the small deposition rate an effect of various nucleation times and an influence of a flux gradient should probably considered here as well. 

The field emission properties (turn-on fields, field amplification) were clearly correlated to the diameter of the tubular C-N structures. The minimal fields and 
highest amplifications were found around $d$ = 4 - 6 mm, i.e., where the lowest overall values for the tube diameter are obtained ($\sim$50 nm). Field emission is a highly selective process, and only structures showing the highest field amplification (e.g., the smallest tube diameter) will participate in a significant proportion to the electron emission. The fact that the emission is not as efficient for larger diameters shows however that the field emission is not governed solely by the diameter of the smallest tubes. In fact, if this were the case the emission fields should decrease further for larger $d$, as the diameter of the smaller species decreases further to $\sim$30 nm far away from the filament. One observes however on the SEM micrographs (e.g., Fig. 3(d)) that the smaller tubes are surrounded by thicker and higher tubular structures. In such a case, the field amplification is not only determined by the diameter, but also by the distance to the neighbouring tubes \cite{42,43,50}. The field amplification on a film is therefore influenced both by the geometrical shape of the emitter and by its surroundings, as the presence of nearby objects may screen the applied electric field and provoke a decrease in the effective field amplification. 

The C-N nanotubes obtained by HF-CVD compare quite well with pure carbon nanotubes obtained by the catalytic decomposition of acetylene in a tube flow reactor 
at 720$^{\circ}$C \cite{43}. The latter samples were also structured by microcontact printing using catalyst Fe:Ni inks of different densities with patterns identical to the ones used here. The best nanotube sample had onset and threshold fields of 1.8 and 3.3 V/$\mu$m, with a field amplification of $\beta$ = 1208. Carbon nanotube films with slightly inferior properties, e.g, a film prepared with an ink of lower density (resulting in less dense nanotube patterns), showed properties nearly identical to the best C-N nanotubes observed here. Actually, one carbon nanotube film had onset and threshold fields of 3.6 and 7.0 V/$\mu$m for a field amplification of 850, to be compared with 3.7 and 7.2 V/$\mu$m and $\beta$ = 750. It is hence very probable that the factors governing the field emission are very similar for the pure carbon and C-N nanotubes, and that the incorporated nitrogen plays very little, if any, role in the field emission. 

Finally, we can also compare the patterned C-N tubes with continuous films obtained by the same method \cite{33}. The best emission obtained on a continuous film was 
characterised by onset and threshold fields of 3.8 and 7.8 V/$\mu$m and a $\beta$ = 675, i.e. slightly inferior to the structured sample at $d$ = 5 mm. The continuous films show however poor homogeneity, and the structured deposition of the catalyst allows a far better control over the emission properties of the sample. 

Due to the single filament configuration a strong temperature gradient appears, which give rise to the observed variations in tube diameter. The most promising range for application regarding to the field emission properties was found to be at $d$ = 4 - 6 mm corresponding to T$_{sub}$ = 720-760$^{\circ}$C. However, once the optimal conditions have been determined, homogeneous and controlled film growth can be obtained by adjusting $r$ and T$_{fil}$ in a multiple parallel filament arrangement as described recently \cite{34}.

\section*{Acknowledgements}

The Swiss National Science Foundation (SNSF) is acknowledged for the financial support. The electron microscopy was performed at the Centre Interdepartmental de 
Microscopie Electronique of EPFL. The authors are grateful to Yann von Kaenel from Departement des Materiaux of EPFL for technical assistance in Raman spectroscopy 
and to Laszlo Forro from Department de Physique of EPFL for fruitful discussions.

\end{document}